\documentclass[reprint,aps,prl,superscriptaddress]{revtex4-2}

\usepackage{graphicx}
\usepackage{dcolumn}
\usepackage{bm}
\usepackage{amsmath,amssymb}
\usepackage[colorlinks=true,linkcolor=blue,citecolor=blue,urlcolor=blue]{hyperref}

\DeclareMathOperator{\We}{We}
\DeclareMathOperator{\Bo}{Bo}
\DeclareMathOperator{\Oh}{Oh}

\newcommand{\change}[1]{\textcolor{black}{#1}} 

\usepackage{placeins}

\begin{document}

% Title
\title{Drops can Perpetually Bounce over a Vibrating Wettable Solid}

% Authors and affiliations
\author{Lebo Molefe}\thanks{These authors contributed equally to this work.}
\affiliation{Engineering Mechanics of Soft Interfaces (EMSI) Laboratory, EPFL, Lausanne, Switzerland}
\affiliation{Laboratory of Fluid Mechanics and Instabilities (LFMI), EPFL, Lausanne, Switzerland}
\author{Tomas Fullana$^*$}\email{tomas.fullana@epfl.ch}
\affiliation{Laboratory of Fluid Mechanics and Instabilities (LFMI), EPFL, Lausanne, Switzerland}
\author{François Gallaire}
\affiliation{Laboratory of Fluid Mechanics and Instabilities (LFMI), EPFL, Lausanne, Switzerland}
\author{John M. Kolinski}
\affiliation{Engineering Mechanics of Soft Interfaces (EMSI) Laboratory, EPFL, Lausanne, Switzerland}

\begin{abstract}
On the surface of a vibrating liquid bath, instead of coalescing, a drop will continually bounce on a thin film of air between the drop and the free surface, giving rise to rich chaotic dynamics and quantum analog behavior. However, perpetual bouncing is yet to be demonstrated on a vibrating rigid solid, where the control of the drop's motion is not limited by the bath dynamics. Here we show that vibration of an atomically smooth mica surface prolongs a drop's hovering state by several orders of magnitude, increasing the time to wet from less than a second to several minutes. The excitation of the second spherical harmonic mode determines a transition between a bouncing state with high-amplitude rebounds, and a bound state, where the drop's motion is locked onto the vibrating solid above a thin air layer. We further develop a coupled linear spring model, accounting for the drop's deformation, that predicts bouncing trajectories without fitting parameters. Our results extend the scope of bouncing drop studies beyond the liquid bath and highlight the potential for precision manipulation of small liquid quantities in air.
\end{abstract}

\maketitle

\textit{Introduction—}Reducing liquid-solid contact facilitates the incredible performance of solid surfaces with anti-fouling \cite{preventing_mussel_adhesion_using_lubricant_infused_materials_2017, self_ejection_salts_other_foulants_superhydrophobic_surfaces_enable_sustainable_anti_fouling_2023}, anti-icing~\cite{understanding_effect_superhydrophobic_coatings_energy_reduction_anti_icing_systems_2011, robust_prototypical_anti_icing_coatings_with_self_lubricating_liquid_water_layer_ice_substrate_2013}, antibacterial \cite{anti_biofouling_properties_femtosecond_laser_induced_submicron_topographies_elastomeric_surfaces_2020} and self-cleaning~\cite{purity_sacred_lotus_escape_contamination_biological_surfaces_1997, lotus_effect_wetting_self_cleaning_2016} properties. The Leidenfrost effect \cite{leidenfrost_drops_2003, leidenfrost_dynamics_2013}, where a drop levitates on its own vapor film above a hot surface for an extended period, eliminates wetting completely, but heat could damage sensitive chemistry in the drop, and leads to evaporation of the entire liquid mass. A stable hovering state for a liquid drop on a solid surface at room temperature is yet to be demonstrated. 

In this Letter, we experimentally show that surface vibration can extend a hovering state for a drop on a smooth solid by several orders of magnitude \cite{bouncing_water_drops_2000, drops_bounce_perfectly_hydrophilic_surfaces_2014, wettability_independent_bouncing_flat_surfaces_mediated_thin_air_films_2015, bouncing_thin_air_2015}, demonstrating a purely kinetic analog of the Leidenfrost effect. Drops hover near the solid in a no-contact state, with their bouncing amplitude controlled by the surface vibration amplitude and frequency. These paradoxical gravity-defying states have been observed for vibrating liquid baths \cite{anomalous_phenomena_fluid_under_action_vibration_2008, floating_under_levitating_liquid_2020, liquid_walls_interfaces_arbitrary_directions_stabilized_vibrations_2021, from_bouncing_floating_noncoalescence_drops_fluid_bath_2005, walking_orbiting_droplets_2005, particle_wave_association_fluid_interface_2006}, where the free surface deformation drives the drop's rich dynamics \cite{drops_bouncing_vibrating_bath_2013, role_droplet_deformations_bouncing_droplet_dynamics_2013, drops_walking_vibrating_bath_hydrodynamic_pilot_wave_theory_2013, exotic_states_bouncing_walking_droplets_2013,single_particle_diffraction_interference_macroscopic_scale_2006, unpredictable_tunneling_classical_wave_particle_association_2009, wavelike_statistics_pilot_wave_dynamics_circular_corral_2013, hydrodynamic_analog_friedel_oscillations_2020, emergent_order_hydrodynamic_spin_lattices_2021}. Here, because the solid is rigid, the drop's resonant behavior is governed by its deformation; we find that a simple reduced-order model that couples the drop's center-of-mass and its second harmonic mode deformation is sufficient to predict drop trajectories  without fitting parameters.

\begin{figure*}[t]
    \centering
    \includegraphics[width=\textwidth]{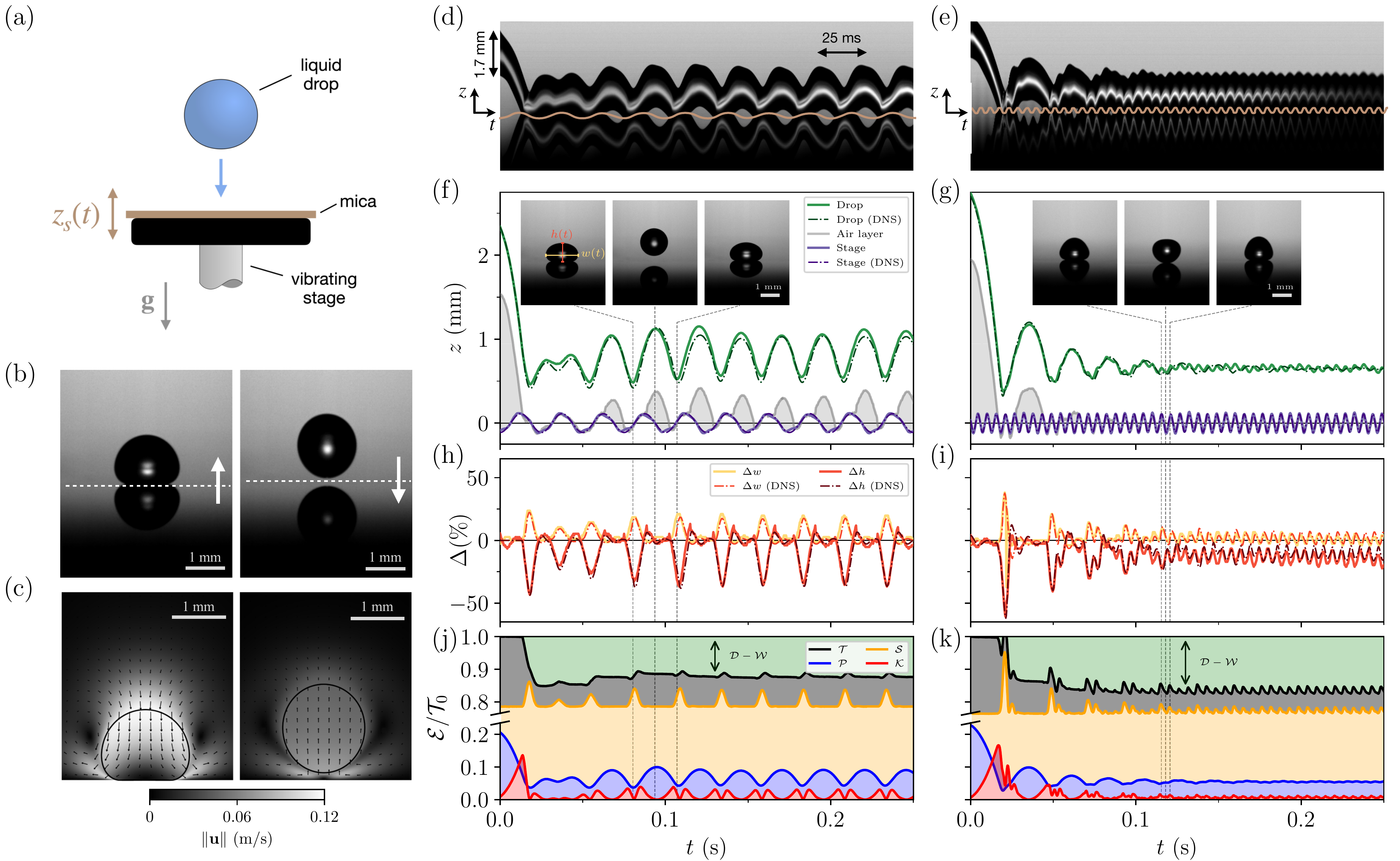}
    \caption{Drop bouncing on air above a vibrating smooth solid. (a) Schematic of the experimental setup. (b) Still experimental images with dashed lines indicating the surface position and arrows indicating its direction of motion. \change{(c) Corresponding snapshots from the DNS.} \change{(d),(e)} Kymographs from a cut through the drop center with the stage position indicated by a solid line; here $r_0 = 0.76$~mm and $A_s = 0.21$~mm. \change{(d)} Bouncing case with $f_s = 40$~Hz ($\Omega \approx 1.2$ and $\Gamma \approx 0.7$). \change{(e)} Bound case with $f_s = 180$~Hz ($\Omega \approx 5.2$ and $\Gamma \approx 14.8$). \change{(f),(g)} Drop center-of-mass position $z(t)$, stage position $z_s(t)$, and air layer thickness for both cases. Insets show image sequences of a cycle, with $\Delta t = 0.014$~s (bouncing case) and $\Delta t = 0.0026$~s (bound case). \change{(h),(i)} Variation of drop's width $\Delta w$ and height $\Delta h$ as a percentage of the drop diameter. \change{(f)--(i)} DNS matching the drops' trajectories and deformations in dash-dotted lines. \change{(j),(k) Normalized drop energies in DNS for the bouncing and bound cases.}}
    \label{fig:bouncing_vs_bound}
\end{figure*}

For a drop impacting a stationary solid surface, the dynamics are governed by the contributions of inertia, gravity, and viscosity relative to surface tension. The three dimensionless groups are the Weber number $\We = \rho r_0 V^2 / \gamma$, the Bond number $\Bo = \rho g r_0^2 / \gamma$, and the Ohnesorge number $\Oh = \eta \, (\gamma \rho r_0)^{-1/2}$ ($\rho$: liquid density, $\gamma$: surface tension, $\eta$: dynamic viscosity, $r_0$: drop radius, $V$: impact velocity, and $g$: gravitational acceleration). In our experiments, surface tension is a significant restoring force~\cite{bouncing_water_drops_2000, when_does_impacting_drop_stop_bouncing_2023}, which allows bouncing \change{without contact} on initial impact due to the presence of a thin compressed air film between the drop and the surface~\cite{skating_film_air_drops_impacting_surface_2012, drops_bounce_perfectly_hydrophilic_surfaces_2014, wettability_independent_bouncing_flat_surfaces_mediated_thin_air_films_2015, bouncing_thin_air_2015, characteristic_rupture_height_mediating_air_film_beneath_impacting_drop_atomically_smooth_mica_2023, sprittles_gas_2024}. In the case of a vibrating surface with position given by $z_s(t) = \frac{1}{2}A_s \sin(\omega_s t + \phi_s)$ [Fig.~\ref{fig:bouncing_vs_bound}(a)], the vibration frequency $f_s = \omega_s/(2\pi)$ and amplitude $A_s$ control the drop's rebound dynamics via two additional parameters: the vibration number $\Omega = \omega_s / \omega_d$ and the dimensionless acceleration $\Gamma = \frac{1}{2} A_s \omega_s^2 / g$, where $\omega_d = (\gamma / \rho \, r_0^3)^{1/2}$
is the drop’s Rayleigh frequency~\cite{capillary_phenomena_jets_1879}.

\textit{Generation of bouncing drops—}We use a high-speed camera to measure the dynamics of bouncing silicone oil drops with $\rho = 941 \pm 9$~kg/m$^3$, $\eta = 20.17 \pm 0.07$~mPa.s, $\gamma = 21.2 \pm 0.2$~mN/m and $r_0 = 0.78 \pm 0.05$~mm. A smooth mica sheet, with roughness on the order of a few nanometers (see Supplemental Material~\cite{supp_mat})\nocite{silicone_oil_datasheet_2011,molefe_coating_2024,daerr_pendent_drop_2016,ducker_measurement_1992,popinet_gerris_2003,afkhami_height_2008,afkhami_height_2009,scardovelli_direct_1999,popinet_front-tracking_1999,tryggvason_direct_2011,brackbill_continuum_1992,bell_second-order_1989,zhang_impact_2022,sharma_energetics_2020}, is attached to the vibrating stage and the drop is released from a syringe located between $z_0 = 2-4$~mm above the surface depicted in Fig.~\ref{fig:bouncing_vs_bound}(a), off of which it will bounce [Fig.~\ref{fig:bouncing_vs_bound}(b)]. \change{The corresponding dimensionless numbers are $\We \simeq 1.25 - 2.95$ with $V \simeq 0.20 - 0.28$~m/s, $\Bo \simeq 0.26$, and $\Oh \simeq 0.16$, which indicate that the drop will initially bounce off the stage even in the absence of surface vibration~\cite{when_does_impacting_drop_stop_bouncing_2023}.} The stage position $z_s(t)$, driven with a function generator and a mechanical shaker, has amplitude $A_s = 0.15-0.25$~mm, frequency $f_s = 20-180$~Hz, and an arbitrary phase shift at release $\phi_s$.

On the vibrating surface, the drop's time to wet $t_w$ -- the time from initial drop impact to contact initiation -- is extended by over two orders of magnitude. Compared to a typical \change{time to wet} of $t_w \sim 1$~s for a stationary surface~\cite{bouncing_water_drops_2000, wettability_independent_bouncing_flat_surfaces_mediated_thin_air_films_2015, bouncing_thin_air_2015}, we observe bouncing for several minutes (\change{$t_w \sim 300$~s}, Movie S1). In experiments, this prolongation is only limited by the drop's lateral motion, as the drop may encounter occasional defects while roaming the mica surface~\cite{characteristic_rupture_height_mediating_air_film_beneath_impacting_drop_atomically_smooth_mica_2023}. Otherwise, our direct numerical simulations (DNS), carried out with the free software Basilisk~\cite{accurate_adaptive_solver_surface_tension_driven_interfacial_flows_2009, quadtree_adaptive_multigrid_solver_serre_green_naghdi_equations_2015} (see Supplemental Material~\cite{supp_mat}), suggest that a drop retains enough kinetic energy to bounce for an extended period, and possibly indefinitely, as the energy injected through the air layer balances the viscous dissipation.

\begin{figure}[t]
    \centering\includegraphics[width=0.45\textwidth]{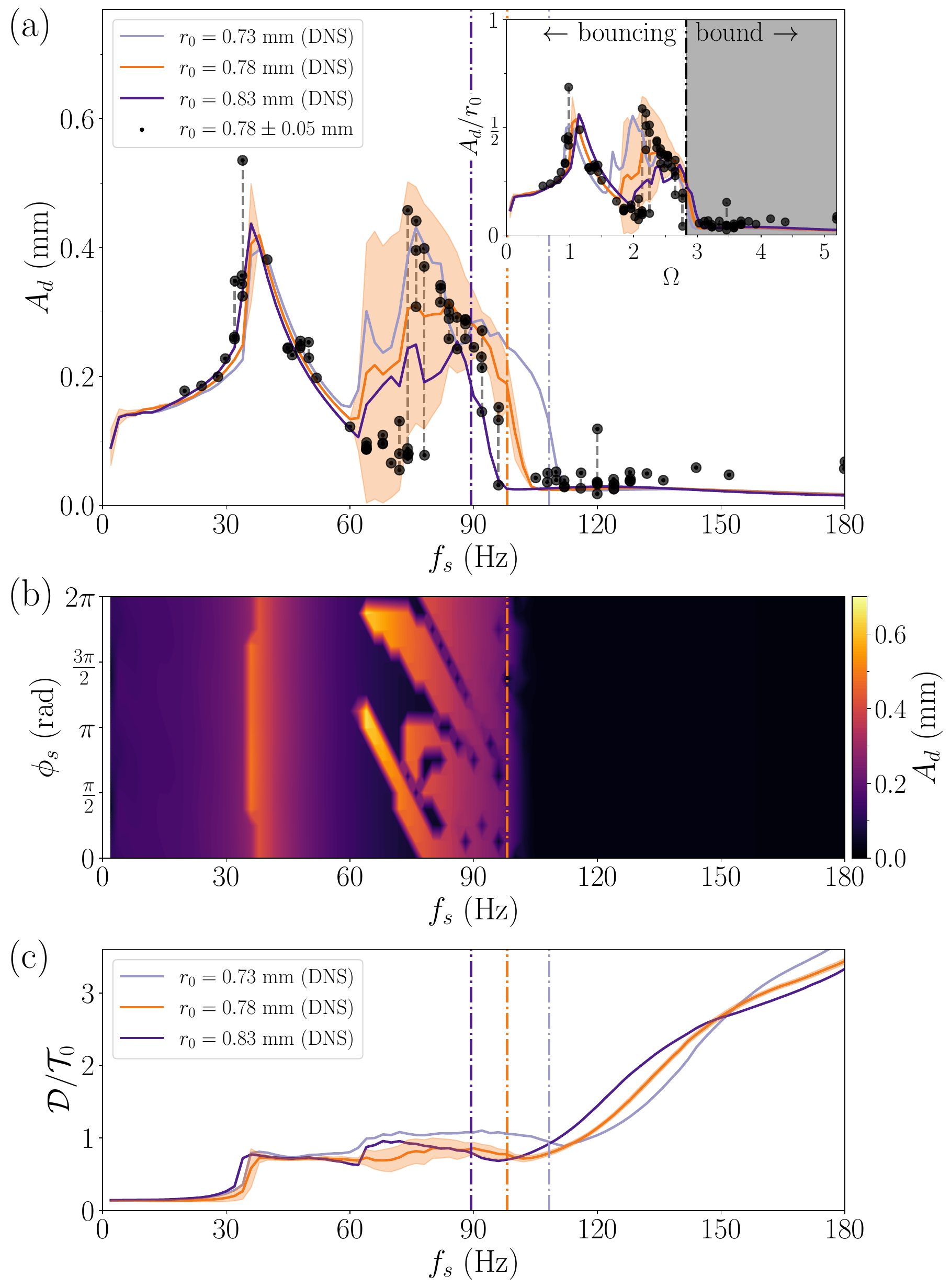}
    \caption{Resonance curve and the bouncing-to-bound transition. (a) Mean drop amplitude $A_d$ for a fixed stage amplitude $A_s = 0.21$~mm. Experimental results with $r_0 = 0.78 \pm 0.05$~mm (black points) compared to phase-averaged DNS results with $r_0 = \{0.73, \, 0.78, \, 0.83\}$~mm (gray, orange and black lines)\change{, for a numerical time span of 0.5 s}. Proposed bouncing-to-bound cut-off frequencies $f_c \approx \{89, \, 98, \, 108\}$~Hz (dash-dotted colored lines) for each radius in the DNS collapse to a single cut-off at $\Omega_c = \sqrt{8}$ in the dimensionless plot (\change{black dash-dotted line in} inset). Experimental spread, shown by dashed vertical lines, falls within one standard deviation of the numerical average (orange shaded band), and is attributable to variation in $\phi_s$. (b) Influence of $\phi_s$ on $A_d$ in the DNS for the same $A_s$ and varied $f_s$ \change{with $r_0 = 0.78$~mm}. \change{(c) Normalized integrated dissipation $\mathcal{D}/\mathcal{T}_0$ for the same DNS cases as in (a).}}
    \label{fig:resonance_curve}
\end{figure}

\textit{Dynamical regimes—}Typical trajectories depicting the motion of the drop's center-of-mass and the air gap show distinct dynamical states for the same stage amplitude, but with different driving frequencies \change{(Movies S2 to S5)}. 
At intermediate frequencies ($f_s = 40$~Hz, $A_s = 0.21$~mm, $\Omega \approx 1.2$, $\Gamma \approx 0.7$), the drop adopts a `bouncing' state with high-amplitude rebounds, leaving a visible air gap during each bounce, as shown in the kymograph in Fig.~\ref{fig:bouncing_vs_bound}\change{(d)}, and images in Fig.~\ref{fig:bouncing_vs_bound}\change{(f)}, inset. By contrast, at higher frequencies ($f_s = 180$~Hz, $A_s = 0.21$~mm, $\Omega \approx 5.2$, $\Gamma \approx 14.8$), the center-of-mass is hardly displaced, indicating that only a thin air film remains and the drop appears dynamically `bound' to the stage [Fig.~\ref{fig:bouncing_vs_bound}\change{(e)} and Fig.~\ref{fig:bouncing_vs_bound}\change{(g)}, inset]. In this bound state, the drop's trajectory and deformation [Figs.~\ref{fig:bouncing_vs_bound}\change{(g)~and~\ref{fig:bouncing_vs_bound}(i)}] are locked onto the forcing frequency, with the interaction mediated by a thin air film. In both cases, we observe large drop deformations; in particular, significant flattening in the bouncing case [Fig.~\ref{fig:bouncing_vs_bound}\change{(f)}, inset] and alternating triangular shapes in the bound case [Fig.~\ref{fig:bouncing_vs_bound}\change{(g)}, inset] that are accurately captured by the DNS (dash-dotted lines) in Figs.~\ref{fig:bouncing_vs_bound}\change{(f)--\ref{fig:bouncing_vs_bound}(i)}.
\change{Figs.~\ref{fig:bouncing_vs_bound}(j)~and~\ref{fig:bouncing_vs_bound}(k) show a normalized energy budget for both cases, with the drop's total energy given by $\mathcal{T} = \mathcal{P} + \mathcal{K} + \mathcal{S}$ ($\mathcal{P}$: potential energy, $\mathcal{K}$: kinetic energy, and $\mathcal{S}$: surface energy). At touchdown, the kinetic energy gained during the initial free fall is transferred into surface energy as the drop spreads, and dissipated through the integrated dissipation energy in the liquid $\mathcal{D}$, which must be counterbalanced by the injected stage energy $\mathcal{W}$ for a drop to bounce indefinitely (see Supplemental Material~\cite{supp_mat} for details on energy calculations). We note, in particular, that the dissipation is higher in the bound case than in the bouncing case, suggesting that the drop retains insufficient energy to fully rebound [Figs.~\ref{fig:bouncing_vs_bound}(j--k) and Fig.~\ref{fig:resonance_curve}(c)].}

\begin{figure}[t]
    \centering
    \includegraphics[width=0.45\textwidth]{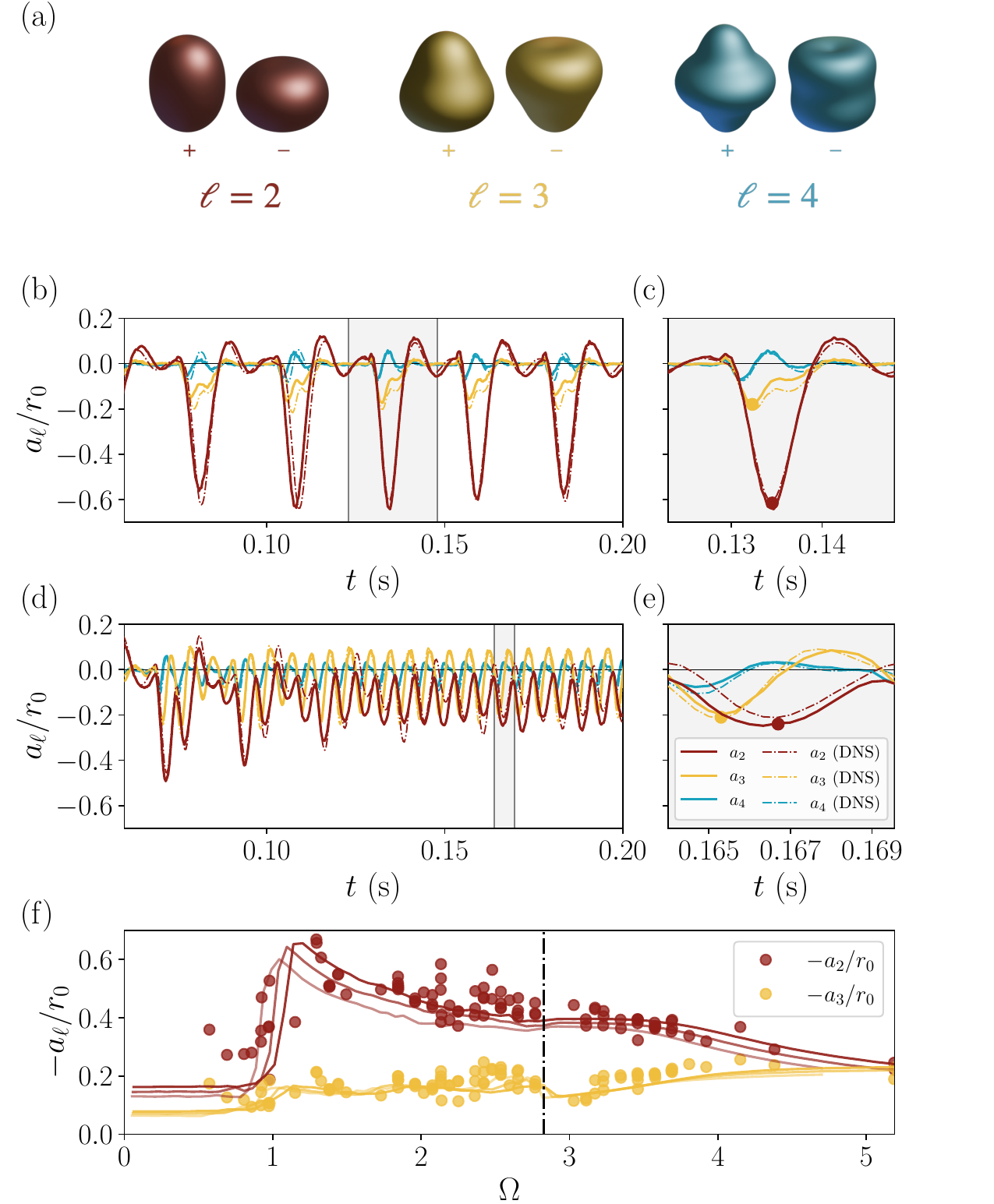}
    \caption{Spherical harmonic decomposition of the drop deformation. (a) Illustration of axisymmetric spherical harmonic modes $\ell= \{2, \, 3, \, 4\}$ with amplitudes \change{$a_\ell/r_0 = \pm 0.3$}. (b) Spherical harmonic mode amplitudes of the bouncing case [Fig.~\ref{fig:bouncing_vs_bound}(c)] with each line color corresponding to one mode  illustrated in (a). (c) Zoom on a single cycle denoted by \change{the shaded region} in (b) where the peak negative values of $a_{\ell = 2, \, 3}$ are marked by a point \change{(minimal values over a rebound period)}. (d),(e) Identical description of the bound case [Fig.~\ref{fig:bouncing_vs_bound}(d)]. (f) Dimensionless peak amplitudes $-a_{\ell = 2, 3}/r_0$ as a function of $\Omega$ \change{with the cutoff frequency $\Omega_c$ as a black dash-dotted line}. DNS results are plotted as dash-dotted lines in (b)--(e) and as solid lines in (f) with $r_0 = \{0.73, \, 0.78, \, 0.83\}$~mm.}
    \label{fig:spherical_harmonic_modes}
\end{figure}

\begin{figure*}[t]
    \centering
    \includegraphics[width=1\textwidth]{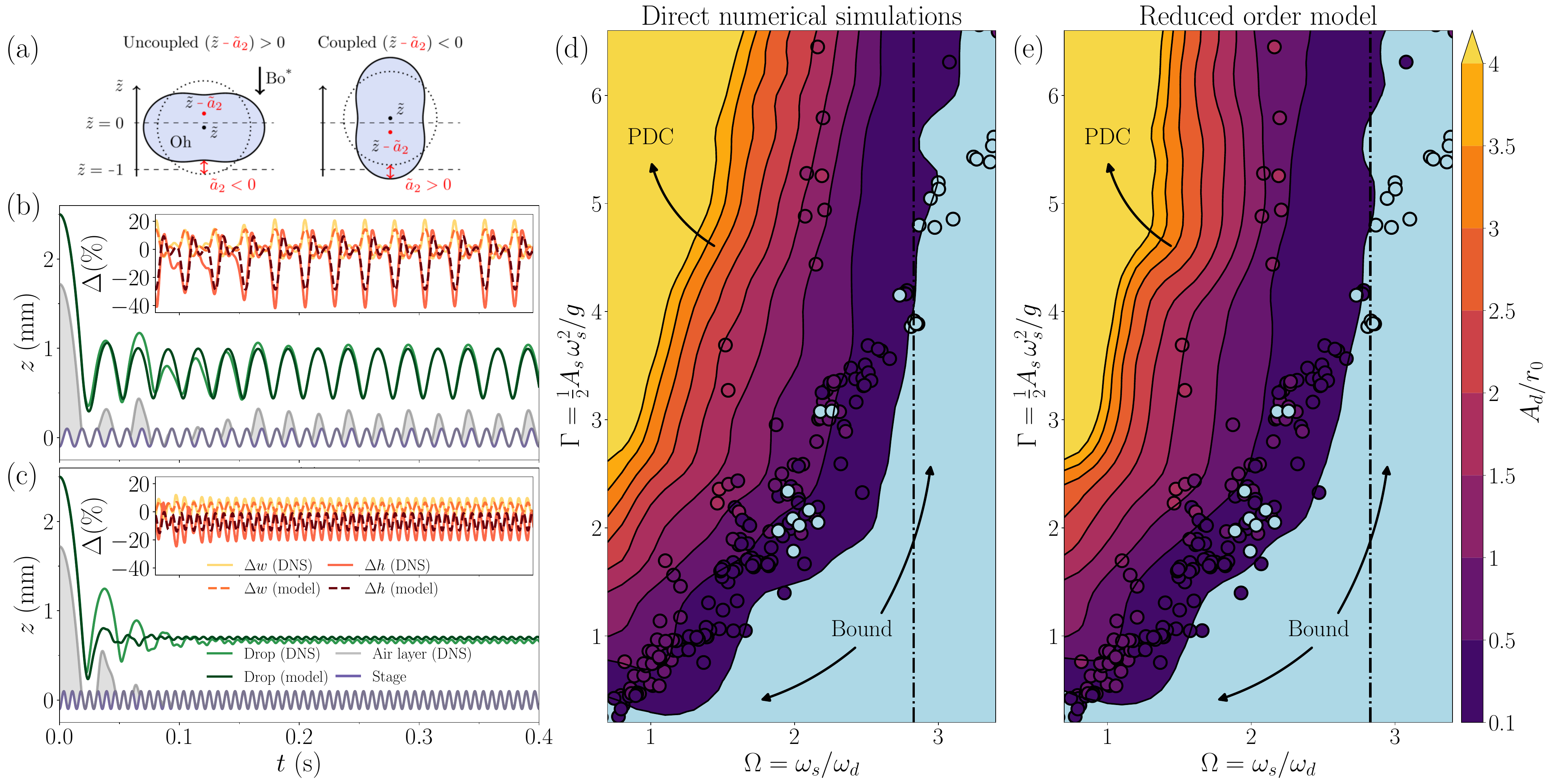}
    \caption{Predictive model captures bouncing and bound dynamics. (a) Schematic of coupled linear spring model defined in Eq.~\eqref{eq:model}. (b),(c) Comparison of the DNS drop center-of-mass position $z(t)$ (\change{light} green line) with the model (\change{dark} green line) in a bouncing case with $f_s = 80$~Hz (b) and in a bound case with $f_s = 140$~Hz (c). Insets show the comparison of DNS and model for the drop's width and height variation as a percentage of the original drop diameter. (d) DNS $\Omega$--$\Gamma$ phase diagram showing the dimensionless drop mean amplitude $ A_d / r_0$ with $r_0 = 0.78 $~mm, averaged over \change{phase shifts} $\phi_s = n \, \pi /2$ with $n = \{0, 1, 2, 3\}$. Experimental data points are overlaid as circles, with colors indicating measured values of $A_d/r_0$. \change{The cutoff frequency $\Omega_c$ is indicated as a black dash-dotted line.} (e) Identical phase diagram computed with the \change{model Eq. \eqref{eq:model}}, showing strong agreement with both numerical simulations and experimental data.}
    \label{fig:dns_vs_model_examples}
\end{figure*} 

\textit{Resonant behavior—}In light of the two very different dynamical states, we explore a wide range of frequencies to better understand the non-linear response of the system. For a drop radius $r_0 = 0.78 \pm 0.05$~mm with natural frequency $f_d = 35$~Hz and stage amplitude $A_s = 0.21$~mm ($A_s/r_0 \approx 0.3$), we vary the forcing frequency over a range from 20 Hz to 180 Hz corresponding to $0.6 \lesssim \Omega \lesssim 5.2$ and $0.2 \lesssim \Gamma \lesssim 13$.
The drop's root-mean-square bouncing amplitude $A_d$ [Supplemental Material~\cite{supp_mat} Sec. \change{S4}] is shown as a function of $f_s$ for measurements (black points) and numerical simulations (solid lines) in Fig.~\ref{fig:resonance_curve}(a). $A_d$ has a sensitive dependence on the initial impact phase~\cite{spreading_dynamics_droplets_impacting_oscillating_hydrophobic_surfaces_2024} [Fig.~\ref{fig:resonance_curve}(b)]; thus, the bouncing amplitude in the DNS is averaged over phase shifts $\phi_s = n\, \pi /8$ with $n = \{0, 1, \dots, 15\}$. Furthermore, to probe the experimental spread in drop radius [see Fig.~S1], simulation results are shown for three radii $r_0 = \{0.73, \, 0.78, \, 0.83\}$~mm (gray, orange, and black lines). \change{The first peak ($f_s \simeq f_d$, $\Omega \simeq 1$) corresponds to an harmonic response of the system}, and, in the second peak ($f_s \simeq 2\, f_d$, $\Omega \simeq 2$), the experimental spread for a given $f_s$ comes from the phase \change{shift $\phi_s$} at initial impact [Fig.~\ref{fig:resonance_curve}(a)] and lies within one standard deviation of the DNS (orange band). The dependence of $A_d$ on $\phi_s$ persists over longer times \change{[Fig.~S3]}, suggesting that the drop retains a memory of its initial impact. \change{These two coexisting states around the second peak correspond to sub-harmonic and harmonic responses of the system for high and low bouncing amplitudes, respectively [Fig.~S4 and Fig.~S5].}

At higher frequencies, we observe a sharp bouncing-to-bound transition at a critical frequency $f_c \approx 100$~Hz [Figs.~\ref{fig:resonance_curve}(a)~and~\ref{fig:resonance_curve}(b), orange dash-dotted line for $r_0 = 0.78$~mm]\change{, where we have defined the bound state by a low drop amplitude $A_d/r_0 < 0.1$.} In addition, the simulations show that $f_c$ depends on drop size, and decreases as $r_0$ increases [Fig.~\ref{fig:resonance_curve}(a), dash-dotted lines]. This behavior suggests that the transition is related to the Rayleigh frequency $\omega_d$, as confirmed by the dimensionless plot [Fig.~\ref{fig:resonance_curve}(a), inset]. \change{Moreover, the total dissipation sharply increases for forcing frequencies above the critical ones $f_s > f_c$ [Fig.~\ref{fig:resonance_curve}(c)], hinting at the excitation of higher order deformation modes in the bound cases.}

\textit{Spherical harmonic decomposition—}We decompose the drop's axisymmetric profile using spherical harmonics, such that
\begin{equation}\label{eq:spherical}
    r(\theta, t) = r_0 + \sum_{\ell = 2}^{\infty} a_\ell(t) \, P_\ell (\cos \theta).
\end{equation}
Here, $\theta$ is the polar angle, $a_\ell(t)$ is the time-varying mode amplitude and $P_\ell(\cdot)$ the Legendre polynomial of the $\ell^{\mathrm{th}}$ axisymmetric spherical harmonic mode. The sum starts with $\ell = 2$ because the $0^{\mathrm{th}}$ mode violates mass conservation, while the $1^{\mathrm{st}}$ mode is a translation (not considered in the center-of-mass frame). In practice, the experimental and numerical data are fitted up to $\ell = 10$, and we find that only the first few modes contribute significantly to the deformation~\cite{bouncing_thin_air_2015} [Fig.~\ref{fig:spherical_harmonic_modes}(a)]. We neglect the air viscosity~\cite{oscillations_fluid_droplet_immersed_another_fluid_1968, free_oscillations_drops_bubbles_initial_value_problem_2006}, and thus the natural angular frequency $\omega_\ell$ associated with each mode is
\begin{equation}\label{eq:mode_frequency}
    \omega_\ell = \sqrt{\ell (\ell - 1) (\ell + 2)} \, \omega_d.
\end{equation}

In the bouncing case [Figs.~\ref{fig:spherical_harmonic_modes}(b)~and~\ref{fig:spherical_harmonic_modes}(c)], the drop spreads horizontally (negative $a_2$) before elongating vertically (positive $a_2$); this leads to an upward translation of the center-of-mass, and \change{to} the lift-off of the drop. By contrast, bound drops [Figs.~\ref{fig:spherical_harmonic_modes}(d)~and~\ref{fig:spherical_harmonic_modes}(e)] deform into alternating triangular shapes with nearly equal amplitudes for modes $\ell = 2 $ and $\ell = 3$, resulting in minimal center-of-mass motion; this prevents escape from the stage. These kinematics reveal that adequate coupling to the $\ell = 2$ mode is necessary for a drop to completely rebound from the surface; $f_2$ defines a critical transition frequency, corresponding to $\Omega_c = \omega_2/\omega_d = \sqrt{8}$. The predicted cutoff value is in remarkable agreement with our experimental and numerical observations, where the bouncing-to-bound transition frequency collapses to the value $\Omega_c$ for all drop sizes [Fig.~\ref{fig:resonance_curve}(a), inset]. Beyond $\Omega_c$, $a_2$ decays whereas $a_3$ increases, corresponding to a limited drop rebound height [Fig.~\ref{fig:spherical_harmonic_modes}(f)].

\textit{Coupled linear spring model—}The pivotal role played by the second spherical harmonic mode motivates the development of a simple coupled linear spring model to predict the drop motion. Whereas most prior models on a liquid bath~\cite{water_spring_model_bouncing_drops_2003, drops_bouncing_vibrating_bath_2013,  role_droplet_deformations_bouncing_droplet_dynamics_2013} neglect the drop deformation, here we couple the \change{dimensionless} drop's center-of-mass $\tilde{z}(t)$ to its shape, characterized by its \change{dimensionless} second deformation mode $\tilde{a}_2(t)$
\begin{equation}
\label{eq:model}
\def\arraystretch{2.2}
\left\{
\begin{array}{c}
\dfrac{\mathrm{d}^2 \tilde{z}}{\mathrm{d} \tilde{t}^2} - H(-\tilde{z}^*) \left[ \operatorname{max}\left\{- \tilde{b}_2 \dfrac{\mathrm{d} \tilde{z}}{\mathrm{d} \tilde{t}} - \tilde{z}^*, \, 0\right\} \right]  = - \Bo^*, \\ 
    \dfrac{\mathrm{d}^2 \tilde{a}_2}{\mathrm{d} \tilde{t}^2} + 2 \, \tilde{b}_2 \dfrac{\mathrm{d} \tilde{a}_2}{\mathrm{d} \tilde{t}} + \tilde{\omega}_2^2 \, \tilde{a}_2  = H(-\tilde{z}^*) \left[ 5 \, (\tilde{\alpha} - \tilde{\alpha}^2/3) \right],
\end{array}
\right.
\end{equation}
\change{with $\tilde{b}_{\ell = 2} = (\ell - 1)(2 \ell + 1) \Oh = 5 \Oh $ the dimensionless dissipation coefficient~\cite{oscillations_fluid_droplet_immersed_another_fluid_1968}, $\tilde{\omega}_{\ell = 2} = \sqrt{\ell(\ell - 1)(\ell + 2)} = \sqrt{8}$ the dimensionless natural frequency associated to the second mode, $\Bo^* = \Bo \, (1 + \Gamma \sin(\Omega\, \tilde{t} + \phi_s))$ the effective Bond number and $\tilde{\alpha} = \mathrm{d}^2 \tilde{z} / \mathrm{d} \tilde{t}^2 + \Bo^*$ the dimensionless instantaneous center-of-mass acceleration.} The Heaviside step function $H(-\tilde{z}^*)$ triggers the coupling when $\tilde{z}^* = \tilde{z} - \tilde{a}_2 < 0$, where $\tilde{z}^*$ is the displaced center-of-mass that accounts for the drop deformation \change{(see appendix for the complete model derivation)}. 

We compare the drop's trajectory and deformation from the DNS to the model without any fitting parameters. \change{In the model, the dimensionless width $\tilde{w} = 2 - \tilde{a}_2$ and height $\tilde{h} = 2 + 2\, \tilde{a}_2$ are extracted using Eq.~\eqref{eq:spherical}.} 
\change{The coupled linear spring} can capture the rich drop dynamics in both the bouncing and bound states [Figs.~\ref{fig:dns_vs_model_examples}(b)~and~\ref{fig:dns_vs_model_examples}(c)]. 
Over a wide range of driving accelerations and frequencies, the drop amplitude $A_d$ predicted from the model agrees nearly exactly with that of the DNS; furthermore superimposed experimental measurements show excellent agreement with both predictions [Figs.~\ref{fig:dns_vs_model_examples}(d)~and~\ref{fig:dns_vs_model_examples}(e)]. Bouncing states exist for accelerations as low as $ \Gamma = 0.3 $, in agreement with similar phase diagrams for drops bouncing on a liquid bath~\cite{drops_bouncing_vibrating_bath_2013}. These results confirm the robustness of the critical frequency $\Omega_c = \omega_2/\omega_d = \sqrt{8}$, above which the drop is bound even for high $\Gamma$ \change{[dash-dotted line in Figs.~\ref{fig:dns_vs_model_examples}(e)~and~\ref{fig:dns_vs_model_examples}(f)]}.

\textit{Control and manipulation of drops—}Our results demonstrate a novel hovering state for a liquid drop on a vibrating smooth solid. We identify two distinct regimes depending on $f_s$ and $A_s$ – a bouncing and a bound state – with a transition governed by excitation of the drop’s second spherical harmonic mode. From this insight, we develop a coupled linear spring model that predicts drop trajectories without fitting parameters.

Prospective drop manipulation in this no-contact state is facilitated by much more rapid transport than above a liquid layer [Fig.~\change{S6}]\change{, which has been similarly observed for larger puddles levitated by airflow \cite{oscillating_star_shaped_drops_levitated_airflow_2013} and Leidenfrost drops above a heated solid \cite{leidenfrost_drops_2003, leidenfrost_dynamics_2013}.} Unlike the Leidenfrost effect, the vibration method does not require high temperatures, allowing for handling of sensitive chemistries. As a proof-of-concept, we demonstrate a pneumatic control method [Fig.~\change{S7}], opening avenues for precision micromanipulation of liquid in air.

\textit{Acknowledgments—}T.F. acknowledges the Swiss Government Excellence Scholarship ESKAS-Nr: 2023.0357.

J.M.K. conceived the study. L.M. conducted the experiments. T.F. performed the numerical simulations. T.F. and F.G. derived the reduced order model. All authors analyzed the data. L.M. and T.F. drafted the manuscript. All authors contributed to the final version of the manuscript.

\textit{Data availability—}The data that support the findings of this article are openly available~\cite{molefe_2025_16875198}. 
\bibliographystyle{apsrev4-2}
\bibliography{refs}
% \newpage
% \onecolumngrid
 \begin{center}
   \textbf{\large \change{End Matter}}
 \end{center}
 \change{\textit{Appendix: Model derivation—} We now include the complete derivation of the coupled linear spring. We non-dimensionalize the lengths by the the drop's radius $r_0$ and time by the characteristic period $\omega_d^{-1}$ and place ourselves in the frame of reference of the vibrating surface with the stage position fixed at $\tilde{z} = -1$ such that an undeformed drop would be acted upon by the stage when its center-of-mass location is negative ($\tilde{z} < 0$) [left panel of Fig.~\ref{fig:dns_vs_model_examples}(a)].}

\change{Away from the stage, the equation of the center-of-mass position is
\begin{equation}\label{eq:com1}
    \dfrac{\mathrm{d}^2 \tilde{z}}{\mathrm{d} \tilde{t}^2} = - \Bo^*,
\end{equation}
and the equation of second mode amplitude is
\begin{equation}\label{eq:shape1}
    \dfrac{\mathrm{d}^2 \tilde{a}_2}{\mathrm{d} \tilde{t}^2} + 2 \, \tilde{b}_2\dfrac{\mathrm{d} \tilde{a}_2}{\mathrm{d} \tilde{t}} + \tilde{\omega}_2^2 \, \tilde{a}_2  = 0.
\end{equation}}

\change{When the stage acts on the drop, we model the center-of-mass equation as a damped linear spring with a reaction force proportional to the displaced drop center-of-mass position $(\tilde{z} - \tilde{a}_2)$~\cite{drops_bouncing_vibrating_bath_2013} and a dissipation term proportional to the drop velocity $\mathrm{d} \tilde{z} / \mathrm{d} \tilde{t}$, with a dissipation coefficient $\,\tilde{b}_2$, as $2\, \mathrm{d}\tilde{z}/\mathrm{d}\tilde{t} \sim \mathrm{d}\tilde{a}_2/\mathrm{d}\tilde{t}$ from geometrical considerations. Enforcing that the reaction force is always positive~\cite{drops_bouncing_vibrating_bath_2013}, the equation of motion reads
\begin{equation} \label{eq:com2}
    \dfrac{\mathrm{d}^2 \tilde{z}}{\mathrm{d} \tilde{t}^2} - \operatorname{max}\left\{-\tilde{b}_2 \, \dfrac{\mathrm{d} \tilde{z}}{\mathrm{d} \tilde{t}} - \left(\tilde{z} - \tilde{a}_2\right), \, 0 \right\}  = -\Bo^*.
\end{equation}}

\change{Considering a quasi-static approximation of drop shapes in contact with a flat solid as derived by Moláček and Bush~\cite[see Eqs. 13 and 42]{molacek_quasi-static_2012}, the second mode amplitude is forced by $f = \tilde{\alpha} \, (2 \ell + 1 / \ell + 1) \, P_2^{'} (1 - \tilde{\alpha} /3)$, with $P_2^{'}(\cdot)$ the derivative of the second Legendre polynomial $P_2(x) = \frac{1}{2} (3 \, x^2 - 1)$. Specifying the expression with $\ell = 2$, we obtain that $f = 5 \, (\tilde{\alpha} - \tilde{\alpha}^2 /3)$, where $\tilde{\alpha} = \mathrm{d}^2 \tilde{z} / \mathrm{d} \tilde{t}^2 + \Bo^*$ is the dimensionless instantaneous center-of-mass acceleration. The equation of motion for the second mode of deformation when forced by the stage reads
\begin{equation} \label{eq:shape2}
    \dfrac{\mathrm{d}^2 \tilde{a}_2}{\mathrm{d} \tilde{t}^2} + 2 \, \tilde{b}_2\dfrac{\mathrm{d} \tilde{a}_2}{\mathrm{d} \tilde{t}} + \tilde{\omega}_2^2 \, \tilde{a}_2  = 5 \, (\tilde{\alpha} - \tilde{\alpha}^2/3).
\end{equation}
The coupling between the center-of-mass and the second mode of deformation is activated when $\tilde{z}^*  = (\tilde{z} - \tilde{a}_2) < 0$. Contrary to the experiments and the DNS, where the drop never contacts the stage, the model allows the deformed drop to intersect the boundary at $\tilde{z} = -1$ [right panel of Fig.~\ref{fig:dns_vs_model_examples}(a)]. Putting together Eqs.~\eqref{eq:com1}--~\eqref{eq:shape2}, we recover the coupled linear spring model defined in Eq.~\eqref{eq:model}.}

\end{document}

% Title
% \title{Bouncing Drops over a Vibrating Atomically Smooth Surface}
\title{Drops can Perpetually Bounce over a Vibrating Wettable Solid \\ -- Supplemental Material --}

% Authors and affiliations
\author{Lebo Molefe}\thanks{These authors contributed equally to this work.}
\affiliation{Engineering Mechanics of Soft Interfaces (EMSI) Laboratory, EPFL, Lausanne, Switzerland}
\affiliation{Laboratory of Fluid Mechanics and Instabilities (LFMI), EPFL, Lausanne, Switzerland}
\author{Tomas Fullana$^*$}\email{tomas.fullana@epfl.ch}
\affiliation{Laboratory of Fluid Mechanics and Instabilities (LFMI), EPFL, Lausanne, Switzerland}
\author{François Gallaire}
\affiliation{Laboratory of Fluid Mechanics and Instabilities (LFMI), EPFL, Lausanne, Switzerland}
\author{John M. Kolinski}
\affiliation{Engineering Mechanics of Soft Interfaces (EMSI) Laboratory, EPFL, Lausanne, Switzerland}

% \date{\today}

\maketitle
\date{March 7, 2025}

\maketitle

\begin{center}
\textbf{CONTENTS}
\end{center}
% \vspace{-1em}
\begin{itemize}
\setlength\itemsep{0.1em}
  \item[] S1 Drop generation and image processing \hfill 1
  \item[] S2 Solid surface characterization \hfill 1
  \item[] S3 Direct numerical simulations \hfill 2
  % \item[] S4 Energy budget calculation \hfill 3
  \item[] S4 Bouncing amplitude and phase dependence \hfill 3
  \item[] S5 Drop mobility on vibrating solid surfaces \hfill 5
  \item[] References \hfill 8
\end{itemize}

\section*{S1 Drop generation and image processing}

\begin{figure}[t]
    \centering
    \includegraphics[width=0.4\linewidth]{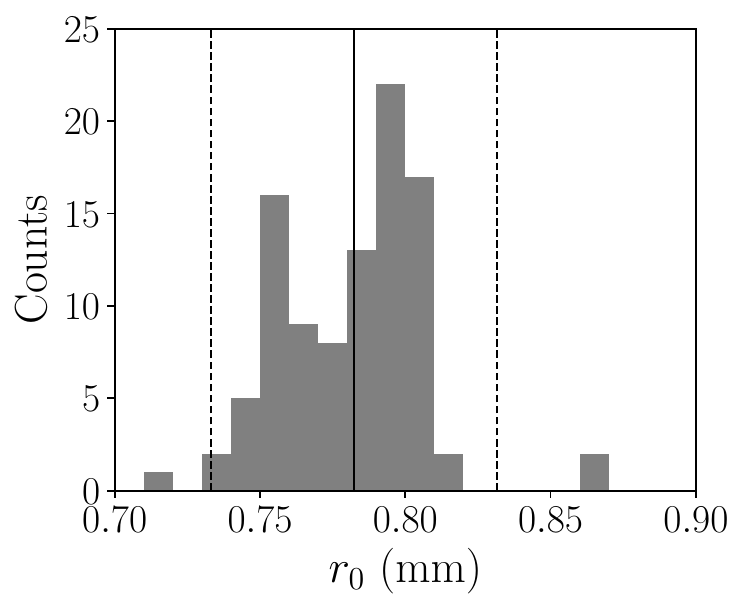}
    \caption{Distribution of drop radius $r_0$ for all experiments. The mean (solid line) is $r_0 = 0.78$~mm, and the margin of error is $0.05$~mm (dashed lines).}
    \label{supfig:drop_radius_histogram}
\end{figure}

The liquid in our dip coating experiments is silicone oil (Bluestar Silicones) \cite{silicone_oil_datasheet_2011}, and we characterized its density $\rho$, viscosity $\eta$, and surface tension $\gamma$ in a previous work \cite{coating_thickness_prediction_viscous_film_rough_plate_2024}. Briefly, density $\rho$ is measured by recording the weight of different liquid volumes between 2--10~mL. Surface tension $\gamma$ is measured by using an ImageJ plugin \cite{pendant_drop_2016} to analyze images of 35 pendant drops. Dynamic viscosity $\eta$ is measured using an Anton Paar MCR 302 parallel plate rheometer: 10 samples are tested and a best fit line is found to relate shear stress to shear strain. Further details and raw data from these measurements may be found in our previous work \cite{coating_thickness_prediction_viscous_film_rough_plate_2024}. Liquid drops of radius $r_0 = 0.78 \pm 0.05$~mm [Fig.~\ref{supfig:drop_radius_histogram}] were produced by a syringe tip having inner diameter $0.15$~mm (Metcal).

To measure the drop's trajectory and deformation, images of bouncing drops are recorded at a frame rate of 3000~fps with a high-speed camera (Photron Nova with IRIX 150 Macro 1:1 lens). Drops are illuminated with an LED (OLSON SSL80 1 PowerStar). To extract the drop position $z(t)$, width $w(t)$, height $h(t)$, and boundary points from the images, we use a threshold to binarize each movie frame to isolate the drop within a manually-identified region of interest. Note that the width $w(t)$ is measured at the drop's widest point, an axis that does not necessarily correspond to the drop's center-of-mass, and the height $h(t)$ is measured at the drop's tallest point. Assuming the drop remains axisymmetric in the experiments, the 2D area centroid is taken to be the center-of-mass position~$z$. Values of $w$ are found by subtracting the difference in position between the rightmost and leftmost pixels of the drop shape; $h$ values were computed similarly with the bottom and top pixels. \change{The air layer thickness in Figs. 1(f)~and~1(g) are characterized by the difference between the drop's lowest point and the stage position.} Finally, to find the drop boundary we locate the edge pixels of the binarized region. To extract the amplitude $a_\ell$ of the $\ell^{\text{th}}$ spherical harmonic mode, we considered a function $r(\theta)$ with origin positioned at the drop's center-of-mass and use least-squares fitting to extract the amplitudes of the modes $\ell = 2-10$. 
% Image processing was performed in Python and code is provided together with an example dataset in the data repository associated with this work.
\section*{S2 Solid surface characterization}

\begin{figure}[t]
    \centering
    \includegraphics[width=0.7\linewidth]{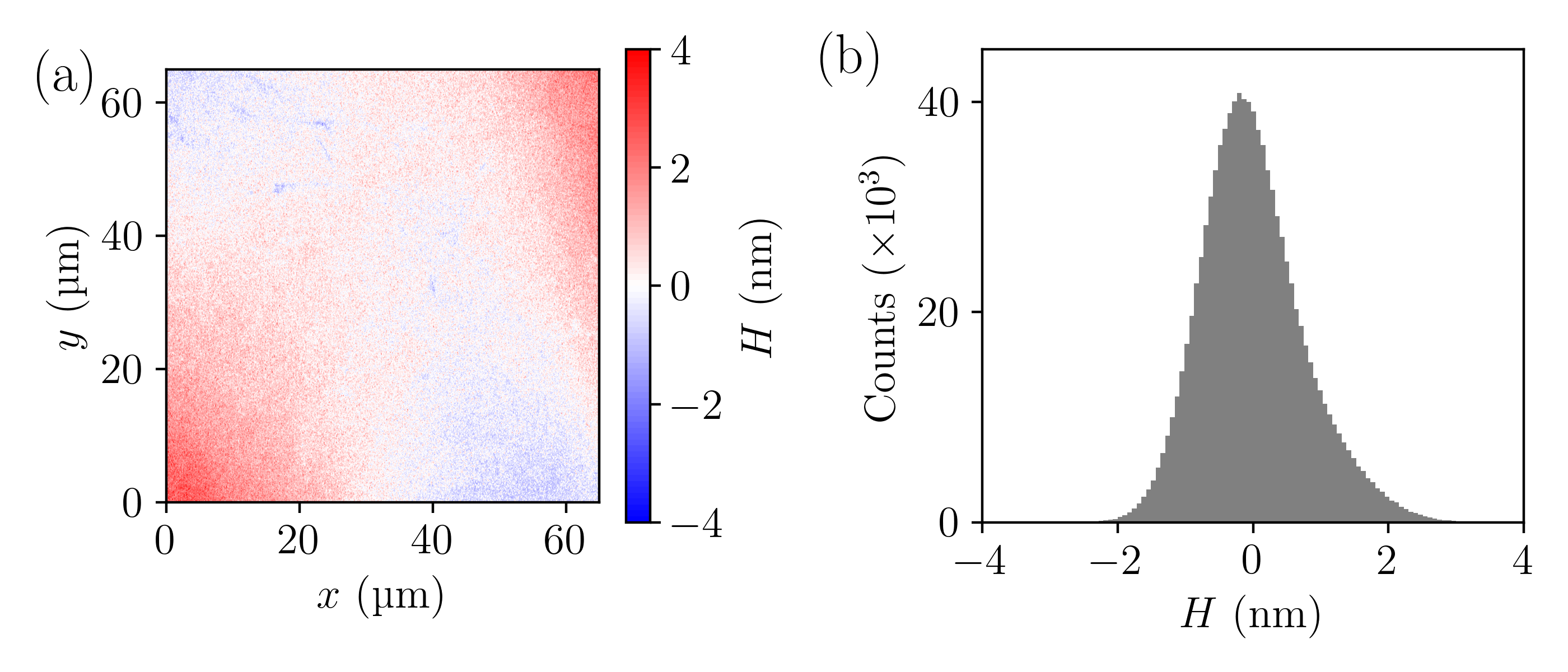}
    \caption{Surface roughness measurement for smooth mica. (a) Surface profile $H(x, y)$ of a freshly-cleaved mica surface obtained with phase shifting interferometric optical profilometry. (b) Histogram of surface profile.}
    \label{supfig:mica}
\end{figure}

We measure the roughness of our mica surfaces using profilometry (Bruker Contour X). A freshly cleaved mica surface is known to be extremely smooth \cite{surface_roughness_plasma_treated_mica_1992}, which is confirmed by our data: the standard deviation of the height profile is $< 1$~nm [Fig.~\ref{supfig:mica}] and the root-mean-square (RMS) roughness is measured to be $\sim3$~nm.

The mica is attached using double-sided tape to a vibration stage on a mechanical shaker, which is driven by a function generator (Keysight 33600A). The vibration stage is carefully leveled with a spirit level. In addition, for several experiments we measure the angle $\theta$ between the line of reflection and the vertical axis of the liquid bridge formed when the drop pinches off from the syringe, finding it to be $\theta = 90 \pm 3^{\circ}$. In this way, we are careful to ensure that the drop's free-fall trajectory -- and thus the direction of gravity -- are perpendicular to the flat surface of the stage. 

To measure the vibrating surface's vertical position, and thus obtain the vibration frequency and amplitude, we firmly affix a small flag-shaped indicator to the side of the vibrating platform. The indicator moves together with the surface, but is offset from the surface position (Movies S2 and S3). To find the offset distance between the indicator and the surface, in one frame of each movie, we locate a line of reflection between the drop flattened against the surface and the drop's mirror image. We measure the distance between the indicator and the reflection line to find the offset and use this constant value to locate the surface position in each movie frame.

\section*{S3 Direct numerical simulations}

Direct numerical simulation results were obtained using the free software Basilisk, a platform for the solution of partial differential equations on adaptive Cartesian meshes~\cite{popinet_gerris_2003, popinet_accurate_2009, popinet_quadtree_adaptive_2015, afkhami_height_2008, afkhami_height_2009}. The flow solver is coupled with the volume-of-fluid method for interface capturing, a technique that is well-known to be suited for solving interfacial flows~\cite{Scardovelli1999,Popinet1999,tryggvason_scardovelli_zaleski_2011}.

We model the liquid and surrounding gas as two incompressible, Newtonian, and non-miscible fluids. We assume that there is no mass and heat transfer at the fluid-fluid interface, and that the surface tension between the two fluids $\gamma$ is constant. The two fluids are denoted $l$ and $g$ for liquid and gas phases, with densities and dynamical viscosities $\rho_l$ and $\rho_g$, $\eta_l$ and $\eta_g$ respectively. We non-dimensionalize the velocity field with the inertio-capillary velocity $V_\gamma = (\gamma / (\rho_l r_0))^{1/2}$ such that the nominal Weber number is unity, pressure with the capillary pressure $P_\gamma = \gamma / r_0$, lengths with the drop radius $r_0$ and time with the inertio-capilary time-scale $\tau = (\rho_l r_0^3 / \gamma)^{1/2}$. The dimensionless, unsteady, incompressible, two-phase Navier-Stokes equations in axisymmetric form where $\tilde{\mathbf{x}} = (\tilde{r}, \tilde{z})$, in the one-fluid formulation and in the frame of reference of the moving stage read 
\begin{align}\label{eq:NS_nd} 
\dfrac{\partial c}{\partial \tilde{t}} + \tilde{\mathbf{u}} \cdot \tilde{\mathbf{\nabla}} c &= 0, \\
\dfrac{\partial }{\partial \tilde{t}} ( \tilde{\rho} \, \tilde{\mathbf{u}}) + \tilde{\mathbf{\nabla}} \cdot (\tilde{\rho} \, \tilde{\mathbf{u}} \, \tilde{\mathbf{u}})  &= -\tilde{\mathbf{\nabla}} \tilde{P}+\tilde{\mathbf{\nabla}} \cdot(2 \Oh \, \tilde{\mathbf{D}})+ \tilde{\mathbf{F}}_\gamma - \Bo^*(\tilde{t}) \cdot \mathbf{e}_z, \\
\tilde{\mathbf{\nabla}} \cdot \tilde{\mathbf{u}} &= 0, 
\end{align}
with $c(\tilde{\mathbf{x}}, \tilde{t})$ the volume fraction equal to $1$ in liquid phase and $0$ in the gas phase, $\tilde{\mathbf{u}} \equiv \tilde{\mathbf{u}}(\tilde{\mathbf{x}}, \tilde{t})$ the flow velocity, $\tilde{\rho} \equiv \tilde{\rho}(\tilde{\mathbf{x}}, \tilde{t})$ the density, $\tilde{P} \equiv \tilde{P}(\tilde{\mathbf{x}}, \tilde{t})$ the pressure, $\Oh \equiv \Oh(\tilde{\mathbf{x}}, \tilde{t})$ the viscosity, $\tilde{\mathbf{D}} = (\tilde{\mathbf{\nabla}} \tilde{\mathbf{u}} + \tilde{\mathbf{\nabla}} \tilde{\mathbf{u}}^T)/2$ the rate of deformation tensor, $\tilde{\mathbf{F}}_\gamma$ the surface tension force~\cite{brackbill_continuum_1992}~and $\Bo^*(\tilde{t}) \cdot \mathbf{e}_z = \Bo \, (1 + \Gamma \sin(\Omega \tilde{t} + \phi_s)) \cdot \mathbf{e}_z$ the body force accounting for the change in reference frame with $\mathbf{e}_z$ the unit vertical vector pointing upwards. In this method, the local density and viscosity are defined from the local volume fraction $c$ 
\begin{align}\label{eq:onefluid}
        \tilde{\rho}(c) &\equiv c + (1 - c) \, \dfrac{\rho_g}{\rho_l}, \\
        \Oh (c) &\equiv \Oh \left(c + (1 - c) \, \dfrac{\eta_g}{\eta_l} \right). 
\end{align}
The equations are approximated using a time-staggered approximate projection method on a Cartesian grid. The advection term is discretized with the explicit and conservative Bell–Colella–Glaz second-order unsplit upwind scheme~\cite{bell_second-order_1989}. For the discretization of the viscous diffusion term, a second-order Crank–Nicholson fully-implicit scheme is used. Spatial discretization is achieved using a quadtree adaptive mesh refinement on collocated grids. A non-wetting boundary condition is enforced by imposing that $c = 0$ at the boundary corresponding to the stage, similarly to previous studies on drop impact within the same numerical framework~\cite{zhang_impact_2022, sanjay_when_2023}.

% \section*{S4 Energy budget calculation}

\change{In this volume-of-fluid framework, the energy budget is computed similarly to~\cite{zhang_impact_2022, sharma_energetics_2020}. The potential energy $\mathcal{P}$, kinetic energy $\mathcal{K}$ and surface energy $\mathcal{S}$ are computed as follows, with $\Omega$ the numerical domain,
\begin{equation}
\begin{array}{c}
\mathcal{P} = \displaystyle \int_{\Omega} \tilde{\rho}_l \, c \, \operatorname{Bo} \, \tilde{z} \, \mathrm{d}\Omega, \\[1em]
\mathcal{K} = \displaystyle \dfrac{1}{2} \int_{\Omega} \tilde{\rho}_l \, c \, \sqrt{\tilde{u}_z^2} \, \mathrm{d}\Omega, \\[1em]
\mathcal{S} = \tilde{\gamma} \, S,
\end{array}
\end{equation}
where $S$ is the surface area of the drop. The energy lost through viscous dissipation $\mathcal{D}$ is computed by integrating the rate of dissipation of mechanical energy per unit mass of the fluid
\begin{equation}
\begin{array}{l}
\displaystyle
    \mathcal{D} = \int_0^{t_{1}} \int_{\Omega} \left[ 
2\Oh(c) \, \left( 
\left( \frac{\partial \tilde{u}_r}{\partial \tilde{r}} \right)^2 
+ \left( \frac{\partial \tilde{u}_z}{\partial \tilde{z}} \right)^2 
+ \left( \frac{\tilde{u}_r}{\tilde{r}} \right)^2 
\right) 
+ \Oh(c) \, \left( 
\frac{\partial \tilde{u}_z}{\partial \tilde{r}} + \frac{\partial \tilde{u}_r}{\partial \tilde{z}} 
\right)^2 
\right] \mathrm{d}\Omega \, \mathrm{d} t.
\end{array}
\end{equation}
Owing to the no-slip condition on the solid surface, the injected stage energy $\mathcal{W}$ is computed as the integrated mechanical power 
\begin{equation}
    \mathcal{W} = \displaystyle \int_0^{t_{1}} \mathbf{\tilde{F}} \cdot \mathbf{\tilde{u}} \, dt = \displaystyle \int_0^{t_{1}} S_c\, \tilde{P} \, \tilde{u}_z \, dt,
\end{equation}
where $S_c$ is the drop's surface area, $\tilde{P}$ is the pressure field and $\tilde{u}_z$ the vertical velocity. These quantities are computed in interfacial cells located one layer above the domain boundary.
}

\section*{S4 Bouncing amplitude and phase dependence}

\begin{figure}[t]
    \centering
    \includegraphics[width=0.8\textwidth]{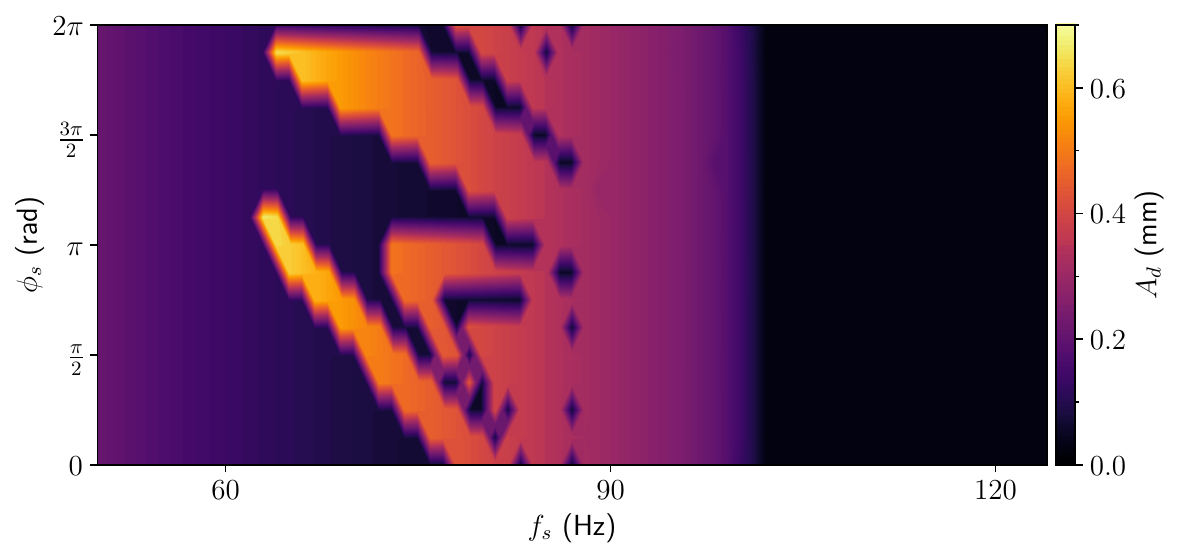}
    \caption{Influence of the impact phase offset $\phi_s$ for longer integration times $t_{1}$. Drop amplitude $A_d$ computed from DNS for varying $f_s$ and \change{$\phi_s = n\, \pi /8$ with $n = \{0, 1, \dots, 15\}$} with $t_{1} = 10$~s.}
    \label{supfig:longertimes}
\end{figure}

\begin{figure}[t]
    \centering
    \includegraphics[width=0.7\textwidth]{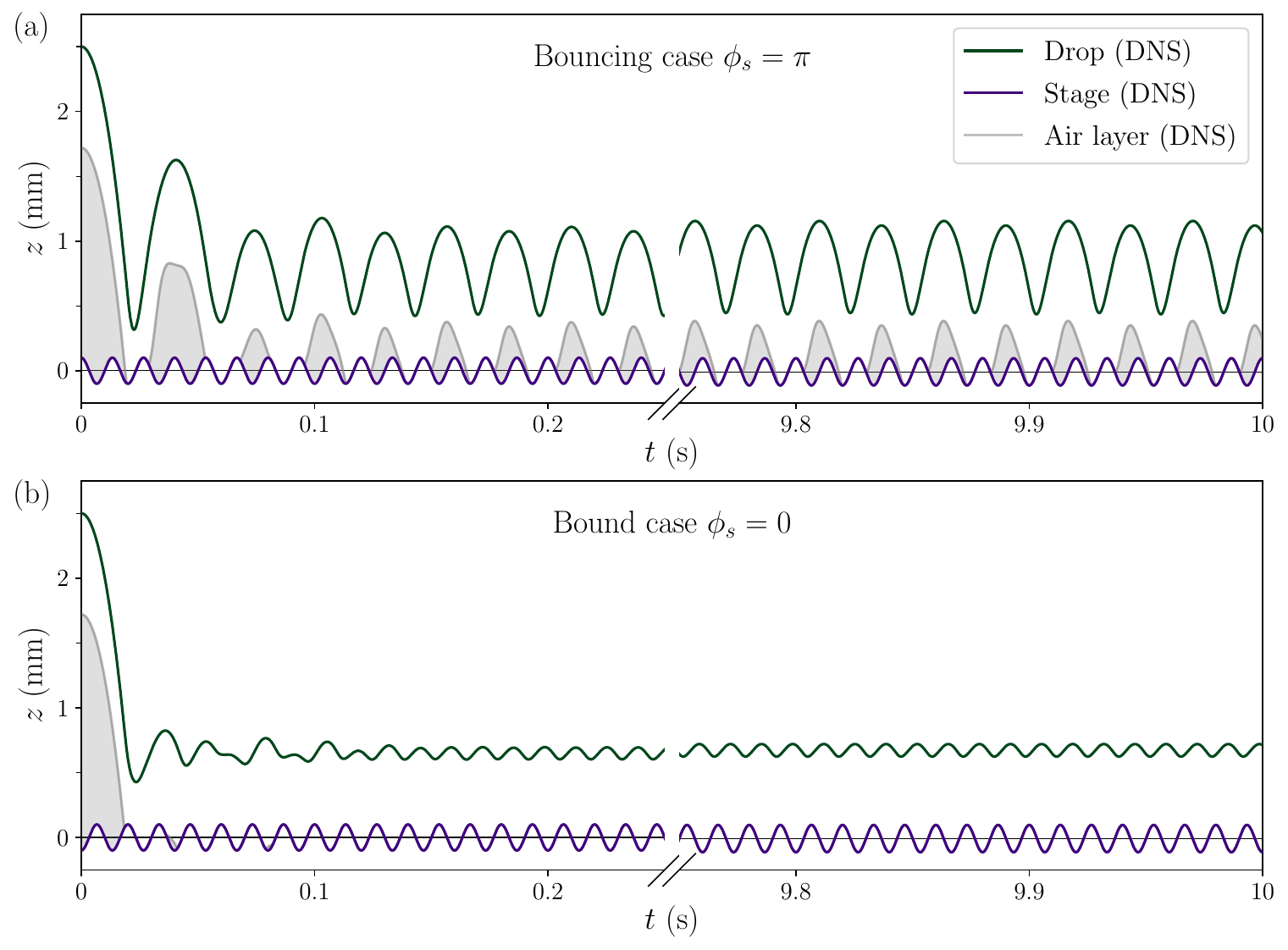}
    \caption{\change{Examples of phase dependence in the DNS at fixed forcing frequency $f_s = 75$~Hz, stage amplitude $A_s = 0.2$~mm and drop radius $r_0=0.78$~mm. (a) Bouncing case for $\phi_s = \pi$. (b) Bound case for $\phi_s = 0$.}}
    \label{supfig:two_states_same_freq}
\end{figure}

The drop's bouncing amplitude $A_d$ is defined as twice the root-mean-square amplitude,
\begin{equation}
    A_d = 2\sqrt{\frac{1}{t_{1} - t_{0}} \int_{t_{0}}^{t_{1}} [z(t) - \langle z \rangle]^2 \,\mathrm{d}t},
\end{equation}
which we use to quantify the peak-to-peak amplitude for the drop's center-of-mass trajectory $z(t)$, where we subtract the mean drop position $\langle z \rangle$ within the integrand, defined by
\begin{equation}
    \langle z \rangle = \frac{1}{t_{1} - t_{0}} \int_{t_{0}}^{t_{1}} z(t) \,\mathrm{d}t,
\end{equation}
and where the integral is performed within a period $t_{0} \leq t \leq t_{1}$ after the initial transient dynamics (we set $t_{0} = 0.1$~s) until the end of the experiment or simulation $t_{1}$. \change{To select a time scale in the simulations that is comparable to the experiments, we set $t_{1} = 0.5$~s in the simulations of the main text. However, we numerically show in Fig.~\ref{supfig:longertimes} that the drop dynamics remain consistent even for longer integration times ($t_{1} = 10$~s). In particular, the drop retains a memory of the initial phase shift $\phi_s$ (Fig. \ref{supfig:longertimes}) and can display two distinct bouncing amplitudes at different phase shifts $\phi_s$ for the same frequency $f_s$ (Fig.~\ref{supfig:two_states_same_freq}).}

\change{In Fig.~\ref{supfig:two_states_same_freq}, we show an example of the dependence on $\phi_s$ for two drop trajectories. In Fig.~\ref{supfig:two_states_same_freq}, we show two time series of drop trajectories $z(t)$ for different impact phase shifts. For $\phi = \pi$ [Fig.~\ref{supfig:two_states_same_freq}(a)], the drop rebounds with a large amplitude and a period equal to twice the forcing period, indicating a sub‐harmonic response at $f_s/2$, corresponding to a bouncing case. In contrast, for $\phi_s = 0$ [Fig.~\ref{supfig:two_states_same_freq}(b)], the drop rebounds with a smaller amplitude once every forcing cycle with an harmonic response at $f_s$, corresponding to a bound case. These examples illustrate how the initial impact phase shift selects between period‐doubled and synchronous bouncing modes.}

\begin{figure}[t]
    \centering
    \includegraphics[width=0.8\textwidth]{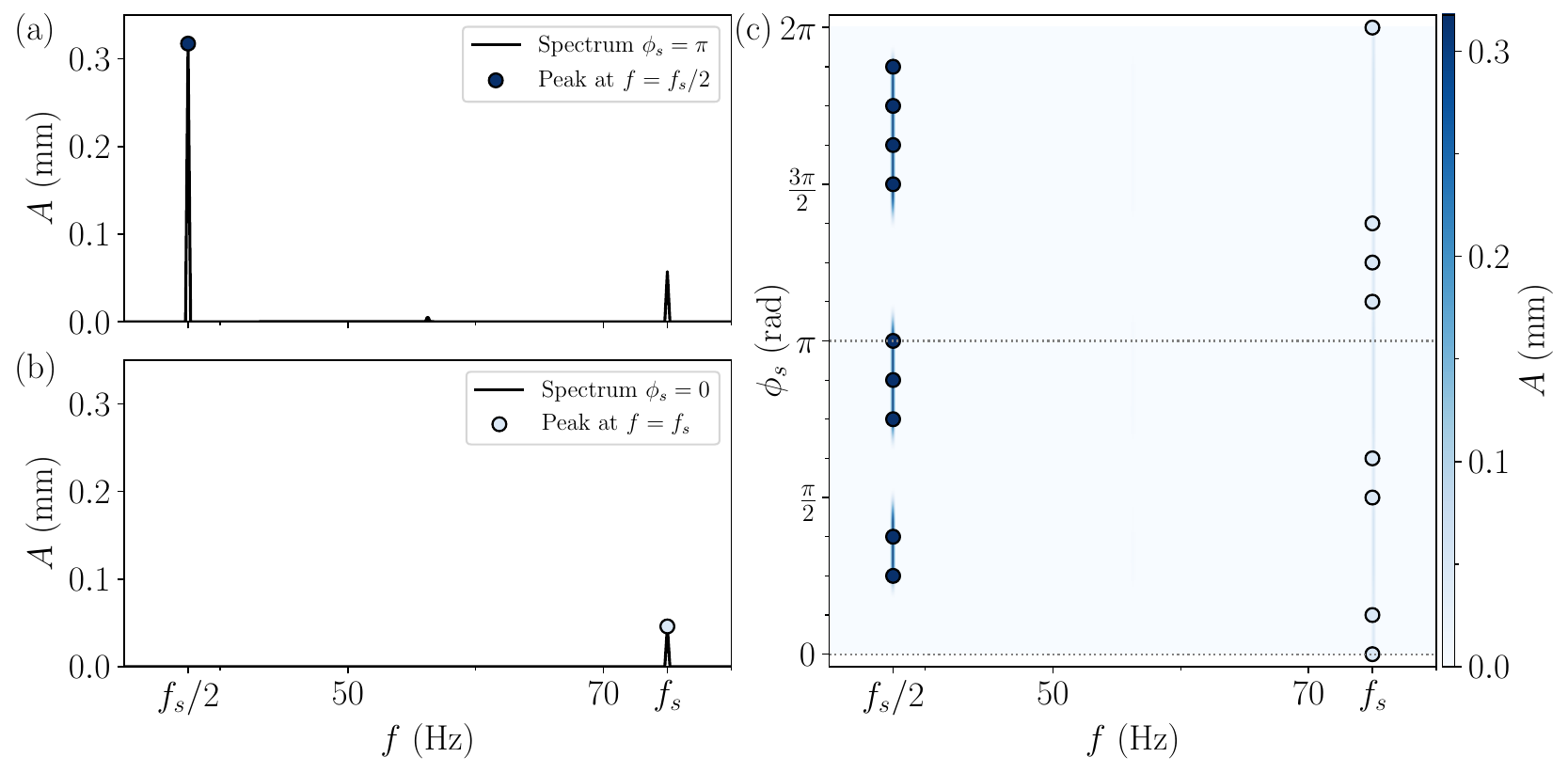}
    \caption{\change{Frequency response of the drop trajectories in the DNS at fixed forcing frequency $f_s = 75$~Hz, stage amplitude $A_s = 0.2$~mm and drop radius $r_0=0.78$~mm. (a) Bouncing case for $\phi_s = \pi$ with sub-harmonic peak at $f = f_s/2 = 37.5$~Hz. (b) Bound case for $\phi_s = 0$ with harmonic peak at $f = f_s = 75$~Hz. (c) Phase diagram of the response amplitude $A$ in the $\phi_s$--$f$ plane for $\phi_s = n\, \pi /8$ with $n = \{0, 1, \dots, 15\}$. The filled points correspond to peak of the spectrum for each phase considered.}}
    \label{supfig:spectrum}
\end{figure}

\change{To further quantify the frequency response of the drop bouncing modes, we compute the discrete Fourier transform of the center-of-mass position $z(t)$ over the last $N = 375$ forcing periods corresponding to $t_N = 5$~s with $f_s = 75$~Hz.  Denoting the uniformly‐sampled signal by $z_n = z(n\Delta t)$ with sampling interval $\Delta t = 3.33 \times 10^{-4}$~s, the complex spectrum is obtained as
\begin{equation}
Z(f_k) \;=\; \sum_{n=0}^{N_s-1} z_n\,e^{-2\pi i\,n\,f_k\,\Delta t}\!,
\end{equation}
where $N_s$ is the total number of samples and $f_k = k/(N_s\Delta t)$.  The one‐sided amplitude spectrum $\,A(f_k)=\frac{2}{N_s}\bigl|Z(f_k)\bigr|$ then gives the amplitude response at each frequency. In Figs.~\ref{supfig:spectrum}(a),(b), we show the spectrum of the two previous examples [Fig.~\ref{supfig:two_states_same_freq}]. As anticipated, the response is seen to be sub-harmonic in the bouncing case [Fig.~\ref{supfig:spectrum}(a)] and harmonic in the bound case [Fig.~\ref{supfig:spectrum}(a)]. The phase diagram of the response over varying phases $\phi_s = n\, \pi /8$ with $n = \{0, 1, \dots, 15\}$ in Fig.~\ref{supfig:spectrum}(c), shows the co-existence of these two stable solutions —period‑doubled or synchronous— and reveals a nonlinear symmetry breaking in the drop’s bouncing motion.}

\FloatBarrier
\section*{S5 Drop mobility on vibrating solid surfaces}

\begin{figure}[t]
    \centering
    \includegraphics[width=0.5\textwidth]{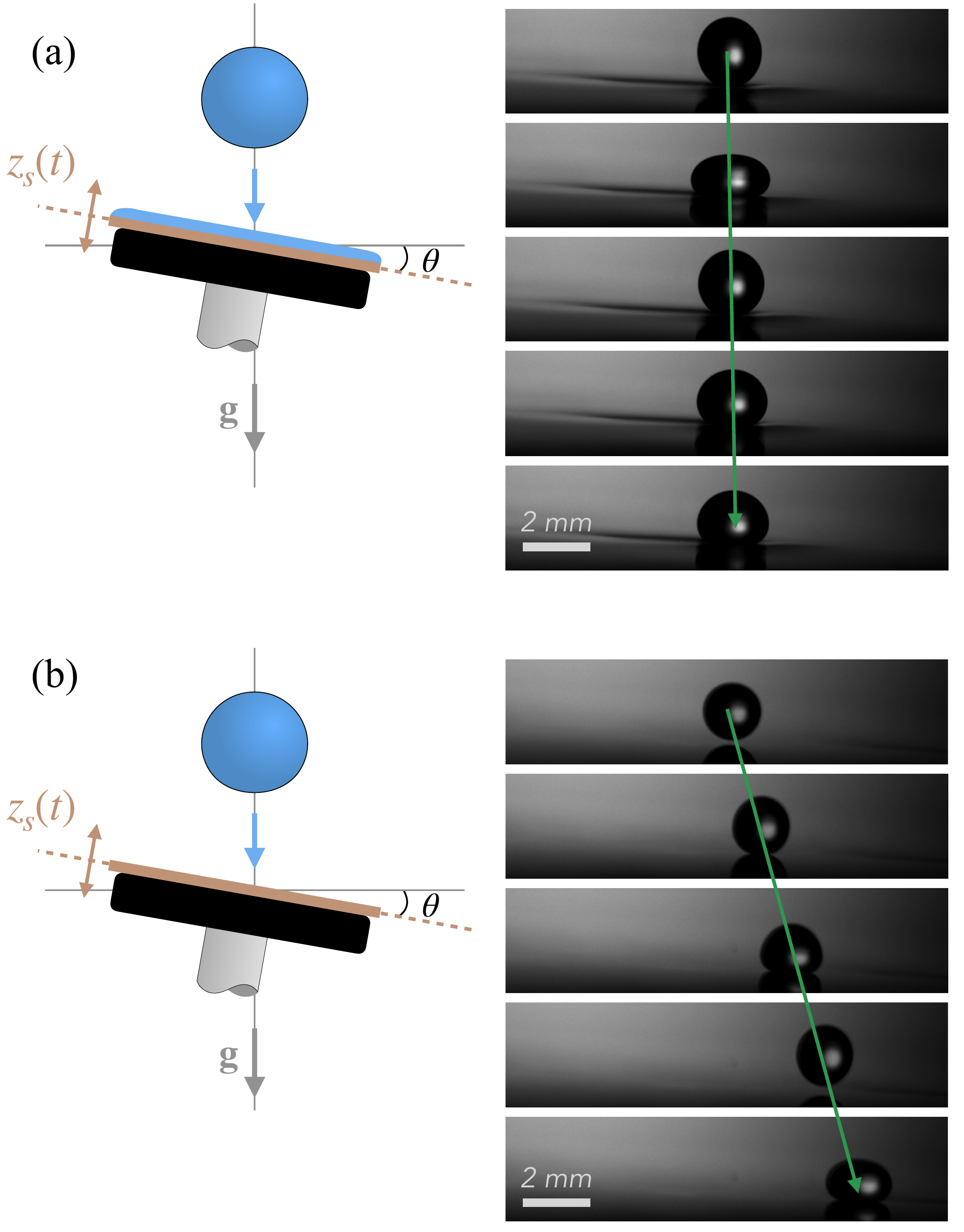}
    \caption{Bouncing drops are highly mobile on solid surfaces. For a minimal horizontal gravitational forcing, imposed by tilting the stage with a given angle $\theta$, (a) a drop bouncing above a vibrating thin liquid layer ($\sim$1~mm thick) has a much lower translation velocity (1.6~mm/s) than (b) a drop bouncing above a vibrating solid surface (80~mm/s). (a),(b) The time interval between images is $\Delta t = 0.0133$~s. The stage frequency is $f_s = 45$~Hz.}
    \label{supfig:highly_mobile_drops}
\end{figure}

\begin{figure}[t]
    \centering
    \includegraphics[width=0.5\textwidth]{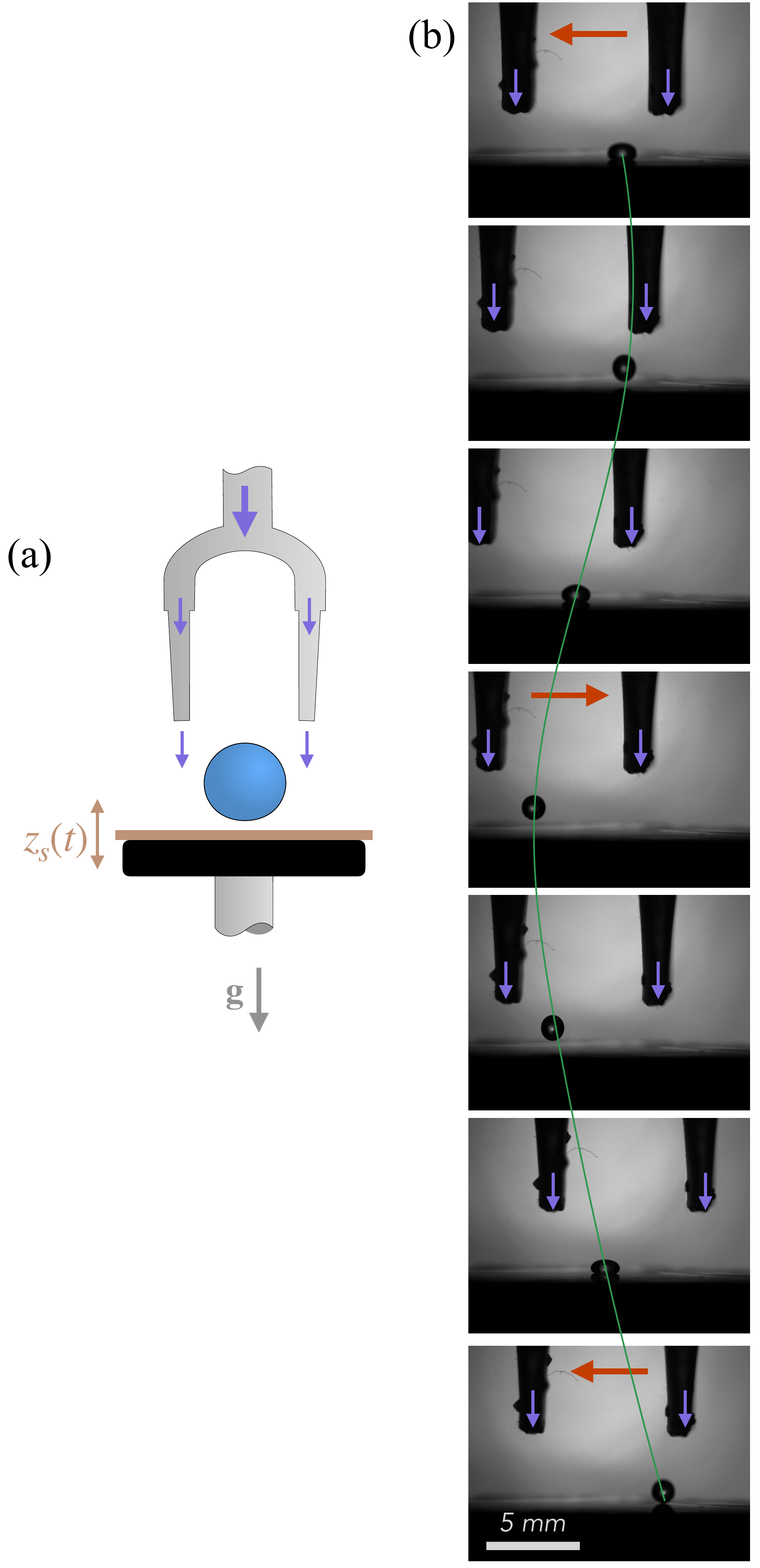}
    \caption{Drop micromanipulation with pneumatic tweezers. (a) Using a pair of symmetric planar jets of air, we control the drop position on the vibrating surface. (b) Demonstration of a left-right translation of the drop's center-of-mass. The time interval between images is $\Delta t = 0.0233$. The stage frequency is $f_s = 40$~Hz.}
    \label{supfig:drop_micromanipulation}
\end{figure}

As a proof-of-concept of the potential manipulation of small liquid quantities in air by vibration of a solid surface, we present here two examples: a comparison of the increased lateral mobility with respect to a vibrating liquid layer [Fig.~\ref{supfig:highly_mobile_drops}] and the micromanipulation with pneumatic tweezers [Fig.~\ref{supfig:drop_micromanipulation}].

%\bibliographystyle{apsrev4-2}
%\bibliography{refs}